\documentclass[reprint,aps,prstper]{revtex4-2}

\usepackage{graphicx}
\usepackage{hyperref}
\usepackage{footnote}
\usepackage{amsmath,amssymb}
\usepackage{multirow}
\usepackage{tabularx}
\usepackage{array}
\usepackage[utf8]{inputenc}
\usepackage{enumitem}
\usepackage{subfig} 

\begin{document} 

\title{Advances in apparent conceptual physics reasoning in GPT-4}

\author{Colin G. West$^1$}
\email[]{colin.west@colorado.edu}
\affiliation{$^{1}$Department of Physics, University of Colorado, Boulder, Colorado 80309, USA}

\date{\today}

\begin{abstract}
ChatGPT is an interface built on a large language model trained on an enormous corpus of human text to emulate human conversation. Despite ChatGPT's lack of any explicit programming regarding the laws of physics, recent work  has demonstrated that the version built on GPT-3.5 could pass an introductory physics course at some nominal level and register something close to a minimal understanding of Newtonian Mechanics on the Force Concept Inventory. This work replicates those results and also demonstrates that the latest version of ChatGPT, built on GPT-4, has reached a much higher mark in the latter context. Indeed, its responses come quite close to perfectly demonstrating expert-level competence, with a few very notable exceptions and limitations. We briefly comment on the implications of this for the future of physics education and pedagogy. 

\end{abstract}

\maketitle


\section{Introduction} 
 ``ChatGPT" is a software application designed to mimic human conversation by generating and responding to text, based on a ``large language model" (LLM). It makes use of two recent advances in the LLM field: the ``Transformer" model~\cite{TransformerOriginal} and ``pretraining,"~\cite{Pretraining} from whence arises ``GPT" (it is a [G]enerative, [P]retrained [T]ransformer model)~\cite{OpenAI}. A great deal has been written about both the methods used to produce the ChatGPT system~\cite{FewShot}, and its striking abilities, which range well beyond ordinary conversation~\cite{lawschool, barexam, wharton, Importance1, importance2, Gerd}. In a recent work, Kortemeyer has shown that a particular version of ChatGPT, based on the language model GPT-3.5, ``would narrowly pass [an introductory physics course] course while exhibiting many of the preconceptions and errors of a beginning learner."~\cite{Gerd}

Ref~\cite{Gerd} tested the GPT-3.5 model which existed in January 2023. Since then, ChatGPT has experienced two major updates: one which significantly improves its mathematical ability~\cite{update_math} and a large update whose details remain proprietary to a new underlying model called GPT-4, which exhibits substantial superiority to the GPT-3.5 model in many respects~\cite{GPT4}.

The prior work tested GPT-3.5 on a wide variety of physics assessment types~\cite{Gerd}, including traditional homework problems, exam problems, and the Force Concept Inventory~\cite{FCI}. The OpenAI team responsible for ChatGPT has already demonstrated that GPT-4 displays superior ability at homework- and exam-type problems by showing that can score a ``4" (66th-84th percentile) on the AP Physics 2 exam, where GPT-3.5 only scored a ``3" (30th-66th percentile~\cite{GPT4}; consistent with Ref~\cite{Gerd}). We propose to complete the comparison by assessing GPT-3.5 and GPT-4 with the FCI. We imagine that Kortemeyer and others are likely pursuing similar work and look forward to comparing our results. 

 We have also studied the performance of ChatGPT with the FCI in a previous work~\cite{mypreprint}, which contains many additional background details on the methods reported here. This project was in progress concurrently with Ref~\cite{Gerd}, but that work preceded us to preprint by several weeks. We did not become aware of their work, however, until the completion of ours. As Ref~\cite{Gerd} notes, ``ChatGPT is probabilistic, which makes particular results in this case study inherently irreproducible." We therefore view this and our prior work as an independent confirmation of their findings regarding GPT-3.5, and now provide a comparable analysis of GPT-4's performance in conceptual introductory physics.  


\section{Methods}{\label{sec:methods}

The FCI is an influential and heavily-studied assessment designed  with an eye towaards distinguishing real conceptual mastery from the kinds of rote memorization, pattern-matching, and algorithmic calculation\cite{CarlKathy, elby, patternmatching} which students sometimes use in order to pass conventional physics tests--and which an AI might be accused of substituting for true expertise.

While the exact details of the text which was used to ``train" ChatGPT are a proprietary secret, it is known that its reading material was largely drawn from the ``Common Crawl" corpus~\cite{commoncrawlfaq}, an open repository of data scraped from text found on the public internet. Common Crawl allows users to query which domains it has indexed; we used this feature to verify that it has not indexed the handful of websites which might most plausibly contain solutions to the FCI (typically, because these sites are paywalled). Because of this, and for additional reasons described in~\cite{mypreprint}, we believe that the FCI is likely \textit{not} in the training text of ChatGPT and hence that its responses have to represent more than regurgitation of something it ``remembers."

\subsection{Modifying the FCI}

There are 30 sequential multiple-choice items in the FCI, with each item containing five choices (four distractors and one unique correct answer or ``key").~\cite{FCI} Its items cover topics like kinematics, projectile motion, free-fall, circular motion, and Newton's laws. Notably, many of its items contain references to figures, but GPT3.5 is designed only to accept text input. And while the GPT-4 model of ChatGPT advertises multimodal capability which might make it possible to prompt the model with a figure, this feature is not publicly available as of this writing.

Of the 20 items with reference to figures, we find that 13 can be modified by adding text that describes the figures without fundamentally altering the task at hand. In doing so, we take care not provide additional clues or context that would make the problem simpler for ChatGPT. For example, item seven involves a steel ball on a rope, swung in a circle and then suddenly cut free. The question asks about the path of the ball after it is released, and the figure supplies  different possible trajectories. One way to describe these options would be, ``tangential to the circle," ``normal to the circle," etc. But this modification threatens to transform the problem into simple word association--``tangential" with ``circular motion." Instead, we transcribe with reference to cardinal directions: \begin{quote}
    Consider a moment in the ball’s motion when the ball is moving north. At that moment, the string breaks near the ball. Which of the following paths would the ball most closely follow after the string breaks?

\begin{enumerate}[label=(\alph*)]
    
    \item It will initially travel north, but will quickly begin to curve to the west
\item It will travel north in a straight line
\item It will travel northeast in a straight line
\item It will initially travel east, but will quickly begin to curve north
\item It will travel east in a straight line

\end{enumerate}

\end{quote}

Seven additional items were excluded from our initial analysis~\cite{mypreprint} because we felt that descriptions of the figure fundamentally altered the character of the problem (for example, describing items 19 and 20 requires specifying the numerical values of the positions of objects at various times, which invites solution by mapping to a calculation rather than conceptual reasoning). However, while these changes may represent an obstacle to direct comparison with student performance, similar changes were made in the version of the FCI given to GPT-3.5 in Ref~\cite{Gerd}. Consequently, it is appropriate to include such questions here to allow comparisons. 

\begin{table}[ht]
\begin{center}
\begin{tabular}{||c | c||} 
 \hline
Type of change & Items\\ [0.5ex] 
 \hline\hline
None & {1, 4, 29, 30}  \\  [1ex] 
 \hline
 Minor text & { 2, 3, 13, 25, 26, 27 }  \\ [1ex] 
 \hline
 Replace fig (minor) & {5, 7, 9, 10, 11,  15, 16, 17 18, 22, 23, 24, 28} \\  [1ex] 
 \hline
Replace fig (major) & {6, 8, 12, 14, 19, 20, 21} \\ [1ex] 
 \hline
\end{tabular}
\end{center}
\caption{Table of items from the FCI and the ways that they were modified for use in this work. ``Replace fig (major)" refers to changes which may have altered the difficulty or nature of the problem, though without fundamentally changing the scenario described in the problem.}
\label{tab:changes}
\end{table}

Six more items without figures received minor text modifications that should not have affected the nature of the physics being tested (For example, references to ``the previous problem," we replaced with simple restatements of the set-up, so that items could be asked about independently if needed). Finally, four of the items were left entirely unchanged. For a complete breakdown of which items were changed in which ways, see table~\ref{tab:changes}. 

\section{Results}\label{sec:results}

For each item, we administered the question with a basic prompt asking for an answer and a brief explanation. Items were administered to ChatGPT using the version of GPT-3.5 which existed after February 20th (notably including its mathematics update) and to the version of GPT-4 which was made public after March 16th. 

\subsection{Answer Choices}

In its first attempt, GPT-3.5 gave a correct answer for fifteen of  thirty, or 50\%. This is just short of the 60\% benchmark suggested by the architects of the FCI to represent when a student has ``barely begun to use Newtonian concepts coherently in their reasoning."~\cite{FCI2}. This is also consistent with the average performance reported in Ref~\cite{Gerd}. 

GPT-4, by contrast, achieved a staggering 28/30 on its first attempt, missing items 19 and 26. While there are clear differences between ``teaching a student" and ``upgrading the underlying language model," it is intriguing if somewhat inapt to characterize this jump in performance is by imagining GPT-3.5 as a student ``before" some form of instruction and GPT-4 as the same student afterwards. Computing the normalized learning gain~\cite{hake_gain} of that hypothetical student yields $g = 86.7\%$, which is more than twice the average effect of a pedagogically-designed, interactive physics course.~\cite{vonkorff} 

One can also compare the performance of these models to typical university physics students, though there are limitations because of the way some FCI questions had to be altered. Using results from our own students from a Fall '18 introductory physics course at a large public R1 university (Figure~\ref{fig:posthist}, we find that GPT-3.5 would fall in the 39th percentile among students at the end of the course, and GPT-4 would fall above the 96th. University student scores from other published studies of FCI performance distributions show similar results.~\cite{planinicFCIdistribution}

\begin{figure}[ht]
    \centering
    \includegraphics[width=0.5\textwidth]{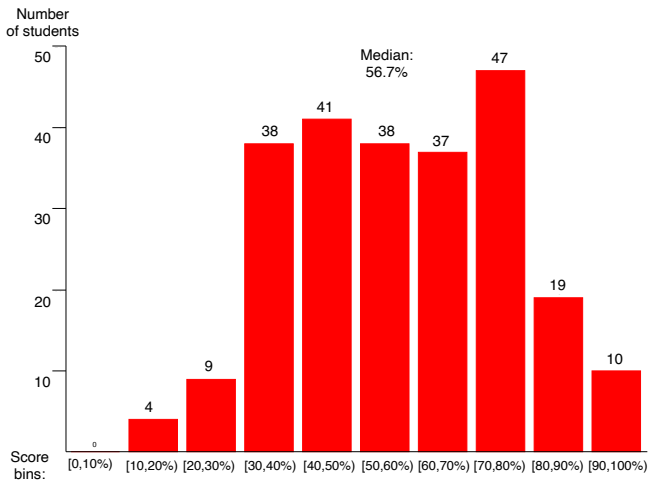}
    \caption{ The distribution of FCI scores at the end of a semester of college physics at a large public R1 university. GPT-3.5's performance of 50\% would put it in the bin containing the median (56\%). GPT-4 falls in the rightmost bin.}
    \label{fig:posthist}
\end{figure}

\subsubsection{Stability}\label{sec:stability}

Because there is a probabilistic element inherent in ChatGPT's operation, we make a brief study of the ``stability" of the models' responses by experimenting with two types of perturbations. First, we use ChatGPT's ``regenerate response" button to simply request that the model start over its process with the same prompt but different random internal conditions. About 1/3 of GPT-3.5's responses displayed significant variation within the first ten requests to regenerate, with the rest differing at most one time in ten. The magnitude of this instability effect is comparable to that found in Ref~\cite{Gerd}. 

We also explore perturbations to the \textit{input}, by repeating a subset of the questions with irrelevant words and sentence structures altered (e.g. ``A boy throws a steel ball straight up" becomes ``a rock is tossed directly upward by a girl). We find that perturbations are much more likely change the responses for the problems GPT-3.5 was already getting completely wrong; correct answers are relatively stable.

By contrast, GPT-4's responses were almost completely insensitive to either type of the perturbation. The median amount of variation in its responses was zero, though some problems displayed variation at the one-in-ten level. Notably, neither of its ``wrong" answers displayed any heightened variability.

\subsubsection{Free Responses}\label{sec:fr}

In our judgement, ten of GPT-3.5's ``free responses," in which it explained the alleged reasoning for its answers, were exactly correct. These responses were typically relatively terse, invoking the correct constellation of concepts, and often making explicit note of facts which ruled out wrong answers as well as the facts which supported the correct one. Another five responses corresponded to the correct answer, but with an ``unforced error" in the explanation--a false statement which was not necessary to support its reasoning in the first place. 

For GPT-3.5's 15 \textit{wrong} answers, the short answer explanations were all strikingly incorrect. Almost all erred in a manner which seemed random or self-inconsistent, unlike the incorrect responses of a human, which will often have an internal logic but proceed from unsound premises. 

GPT-3.5 struggled in particular with problems that depended on reasoning about directions and spatial relationships. Consider item seven: the aforementioned problem regarding a ball released from circular motion. In our wording, the ball is moving north at the time it is released, so the correct answer should be that it travels in a straight line to the north. GPT-3.5 applies some correct concepts but cannot situate them directionally, explaining:  

\begin{quote}
     When the string breaks, the ball will continue to move in a straight line tangential to the circular path it was following at that moment. Since the ball was moving north, the tangential velocity of the ball will be directed to the east.

    However, the force of gravity will also be acting on the ball, pulling it downward towards the Earth. As a result, the ball will follow a curved path towards the west as it falls. 

    Therefore, the correct statement is (A) It will initially travel north, but will quickly begin to curve to the west. 
\end{quote}

Similar errors appeared in many of the problems which made reference to cardinal directions. One possible reason for this is that, while there are good reasons to expect a term like ``tangential" to be preferentially associated in the training corpus with release from circular motion, there is no reason that a particular direction like "north" would be similarly associated. Hence, it can identify the relevant global principle but not identify the relevant local feature.

The contrast with GPT-4 on this problem is useful. It responded:

\begin{quote}
    When the string breaks, the ball will continue in the direction it was moving at the exact moment the string broke, due to its inertia. Since the ball was moving north when the string broke, it will continue to travel north in a straight line.
\end{quote}

Note that this answer was ``stable" and not a lucky guess. Like GPT-3.5, GPT-4 has identified a correct general concept to invoke (inertia), but has also succeeded at drawing a specific inference in this case. We hypothesise that it has found a deeper pattern in its corpus showing how solutions to similar problems select a direction based on the problem statement.  

On the whole, we assess that 26 of GPT-4's responses are entirely correct at the expert level. Three contain small, ``unforced" errors--for example, in item 22, it casually calls the motion of the rocket ``diagonal," despite having correctly identiied the curved trajectory of the rocket in item 21.

Only one response contains the kind of baffling failure of internal logic seen with GPT-3.5. Curiously, it is item 19, also highlighted for GPT-3.5 in Ref~\cite{Gerd}. There, it  displays a mathematical error and a bizarre a logic error. GPT-4 calculates everything correctly, but makes a similar logical misstep at the last moment. 

While a few of the free-response explanations showed minor weaknesses hiding under correct answers, they also occasionally revealed the reverse, as in the case of item 26, which asks about the behavior of a box being pushed against friction. The answer which GPT-4 chooses, "D," would in fact be correct if air resistance were considered non-negligible, and the free-response explanation is consistent with this. When we prompted the model in a new instance after adding the text ``assume air resistance is negligible," it repeatedly gave the correct answer with a correct explanation. 

\subsection{Limitations to GPT-4}\label{sec:limitations}

We note two curious limitations which we observed to the otherwise impressive performance of GPT-4. First, it seems completely incapable of reproducing the reasoning of a novice, even in the face of various prompts asking it to pretend it does not know things like Newton's Laws. During these tests, described in more detail in Ref~\cite{mypreprint}, it resisted changing any of the answers it initially got correct. The picture that emerges is that of a traditional physics professor, who understands the physics well but struggles to recognize their students' difficulties because they can scarcely remember what it was like not to know the concepts by heart. 

Next: in three brief tests we performed, GPT-4 failed spectacularly at responding to questions which did not adopt the canonical tone and nomenclature of a physics problem, even if they related to content it had already seemingly understood on the FCI. For example, inspired by item 2, we prompted the chatbot as follows: 

\begin{quote}
So I was playing with my toddler last night and we were racing his cars and trucks along our dining room table. At one point we started racing them off the edge of the table and my son noticed that the big truck I kept pushing off the side went a lot further before it hit the carpet than when he would push his little sedan off. I couldn't quite tell him why that kept happening but I was guessing it was because the truck toy was so much bulkier. Is that what was going on there?
\end{quote}

The chatbot responded:

\begin{quote}
It seems like your son observed a difference in the behavior of the two toy vehicles when they were pushed off the edge of the table. While there might be several factors at play, one possibility is that the larger, bulkier truck had more momentum than the smaller sedan due to its greater mass.

Momentum is the product of an object's mass and velocity. When you push the toy vehicles off the table, you apply a force to them, causing them to accelerate. The heavier truck would have a greater mass, and if you apply roughly the same force to both toys, the heavier truck would have more momentum as it leaves the table. This greater momentum would cause the truck to travel further before hitting the carpet.
\end{quote}

We stipulate that this is more difficult question than item two on the FCI, as we are asking the respondent to reason backwards that there could have been a difference in the velocities of the two vehicles. But we feel that an expert physicist would not fail to mention this as a possible cause, and, more to the point, would not misapply the concept of momentum as it is invoked here by ChatGPT.

Though several ``informally presented" prompts produced significant errors from the model in our tests, others have found more reasonable answers depending on variations in the prompt wording~\cite{GerdPersonal}. In general, we suspect both limitations arise from the nature of its training text, and are not be intrinsic to the nature of a pretrained transformer models.

\section{Conclusions}\label{sec:conclusions}

 A recent work looking comprehensively at GPT-4's capabilities across many domains observed that it can solve complex mathematical problems and hold meaningful discussions about advanced mathematics, even as it ``can also make very basic mistakes and occasionally produce incoherent output which may be interpreted as a lack of true understanding" and ultimately ``does not have the capacity required to conduct mathematical research."~\cite{bubecksparks} We find that comparable statements are true about its abilities in conceptual introductory physics. 

 Of course, there are many aspects of true ``understanding" -- such as the ability to engage in metaphor~\cite{metaphor} or make use of multiple representations~\cite{representations}--which differentiate experts and novices and which we do not attempt to test here. But for whatever subset of conceptual understanding is captured by the FCI, GPT-4 is capable of responding like an expert in the vast majority of cases. Hence, while it can be comforting to recall its many remaining limitations, we urge the physics education community to focus instead on the derivative of its abilities. 

 It has been observed elsewhere that "[t]he greatest shortcoming of the human race is our inability to understand the exponential function"~\cite{Bartlett2005}. While history contains faulty predictions about the timeline of AI development in both directions~\cite{AIPredictions}, the current pace and nature of the field suggest it will continue to advance rapidly. Indeed, the self-reinforcing nature of the field as AI models learn to train themselves, coupled with the current exponential growth of computing power, suggests that the field as a whole could advance at an exponential or even superexponential rate~\cite{Superexponential}.

Beyond the first-order inquiries into the impact of LLM technologies in classroom settings, which often center around academic integrity~\cite{cotton_chatting}, GPT-4 raises questions also about what we are teaching our students in these classes, and why. Like an integral table in the age of Maple and Mathematica, much of our current curricula may be misaligned with a future in which physicists solve problems through innovative prompting and dialog with an AI system. Certainly, our students would still need their own intuitive grasp of basic physics concepts, just as one cannot use a calculator without grasping the difference between addition and multiplication. But it calls for a renewed audit of exactly which conceptual and mathematical skills are to form the desired learning outcomes of our courses. 

Ultimately, perhaps what is most intriguing is what GPT-4 may tells us not about the future of our classrooms but about our future conceptions of physics as a discipline. GPT-4 succeeds at physics tasks even though it is only explicitly programmed to emulate human \textit{speech} and not provided directly with any of the formal structures of mathematics or the laws of physics. Perhaps this untethering is really its strength. As Marvin Minsky once observed, our own brains do not naturally take on concepts like mathematics through rigid and axiomatized logic. ``Instead of flimsy links in chains of definitions in the mind, each word we use can activate big webs of different ways to deal of things."~\cite{minsky1982} It is possible that the promise of LLMs as an AI technology is precisely that they, too, can engage in fuzzy, contextual reasoning. The success of GPT-4 on tasks of conceptual physics might then be seen as a victory for semantic inferentialism~\cite{brandom1}, and the proposition that, even in STEM fields, ``[c]oncepts are not `in the head,' nor are they completely `out there' – rather, they reside in the game of giving and asking for reasons."~\cite{noorloos2014} Physicists and educators everywhere should be alert to such possible reconceptualizations of their work, which if the exponential holds, will be arriving sooner than most of us can fully appreciate.

\section*{Acknowledgement}
The author thanks Noah Finkelstein for useful discussions regarding the scope and framing of this work.

\bibliographystyle{unsrt}
\bibliography{references}
\end{document}